\documentclass[draft]{agujournal2019}
\journalname{JGR: Space Physics}
\draftfalse

\usepackage{ragged2e}
\justifying

\usepackage{url}
\usepackage{lineno}
\usepackage{soul}
\usepackage{amsmath}
\usepackage{amssymb}
\usepackage{physics}
\usepackage{fix-cm}
\usepackage{tabularx}
\usepackage{float}
\usepackage{bm}

\begin{document}


\title{The Ginzburg-Landau Model of Magnetospheric Chorus: Instabilities and Mode Condensation}


\authors{Brandon Bonham\affil{1}, Amitava Bhattacharjee\affil{2}}
\affiliation{1}{Department of Physics, Princeton University, Princeton, NJ 08540, USA}
\affiliation{2}{Department of Astrophysical Sciences, Princeton University, Princeton, NJ 08540, USA}
\correspondingauthor{Brandon Bonham}{bbonham@princeton.edu}


\begin{keypoints}
\item A Ginzburg-Landau equation describing magnetospheric chorus was recently derived via insights from the free-electron laser model
\item We investigate the linear stability of single-mode solutions of the Ginzburg-Landau equation
\item We find a band of stable modes centered about the frequency with the highest linear growth rate and condensation to stable modes
\end{keypoints}


\begin{abstract}
The analogy between free-electron lasers (FELs) - laboratory devices which generate intense coherent light with tunable frequencies - and whistler wave-particle interactions in the magnetosphere has recently been extended to account for waves with spatially dependent amplitudes and a spectrum of frequencies. The whistler was found to be governed by one of the most well-studied nonlinear equations in physics, the Ginzburg-Landau equation (GLE), which can be used to predict the complex nonlinear physics of multi-mode interactions. In this study, we focus on the single-mode solutions of the GLE and investigate their propagation and stability in the context of magnetospheric chorus. As with FELs, there are two types of instabilities, the Benjamin-Feir instability, where all single modes are unstable, and the Eckhaus instability, where there is a band of stable modes, but all modes outside of the band are unstable. Both stability conditions are given by well known inequalities in the GLE literature. For whistler-mode chorus, we analytically reduce the inequalities to simple expressions and show that, to the extent that the GLE represents magnetospheric chorus, it is Benjamin-Feir stable. We also derive the width of the Eckhaus stability band. We find that the predicted bandwidth is consistent with in situ satellite observations and support our analytical calculations with numerical simulations of the GLE. Our simulations demonstrate the robustness of the stable modes, the evolution from unstable modes to stable ones, and the tendency for mode condensation, whereby a noisy spectrum of modes tends to relax to a single stable mode.  
\end{abstract}


\section{Introduction}

The amplification of magnetospheric chorus by radiation belt electrons can be modeled as a free-electron laser (FEL) - a laboratory device designed to emit intense coherent radiation by injecting a beam of relativistic electrons into a static sinusoidal magnetic field \cite{margaritondo_simplified_2011}. This model was first proposed by \citeA{soto-chavez_chorus_2012}, assuming a single-frequency mode with a spatially independent amplitude. In that work, \citeauthor{soto-chavez_chorus_2012} derived a set of 2$N$+2 nonlinear equations (where $N$ is the number of resonant electrons) for the interaction of whistler waves with electrons, cast in the form of FEL equations. Using a collective variable approach \cite{bonifacio_collective_1986}, \citeauthor{soto-chavez_chorus_2012} reduced this large number of equations to a closed set of just three linear equations, providing an analytically tractable model for chorus amplification. From these linear collective variable equations they made predictions for the amplitude saturation value and timescale which are consistent with in situ satellite observations. 

More recently, \citeA{bonham_whistler_2025} derived the nonlinear collective variable equations for the FEL model, yielding a more precise prediction of the saturation amplitude and timescale, and post-saturation behavior of the whistler and the resonant electrons. Additionally, using a method outlined in the FEL literature \cite{ng_ginzburg-landau_1998}, it was shown that if one allows for spatial variations in the amplitude and a spectrum of wave-frequencies, then the amplitude and phase behavior of the whistler can be described by a cubic-complex GLE \cite{aranson_world_2002,garcia-morales_complex_2012}. Therefore, one can model the rich nonlinear dynamics of a chorus wave packet in space and time using a single partial differential equation, rather than solving a system of 2$N$+2 coupled equations. 

Among other interesting predictions, such as the existence of solitary waves and periodic amplitude spikes, the GLE also admits single-mode solutions and the possibility of mode condensation. Mode condensation is characterized by the collapsing of power from an initially broad spectrum of modes down to a narrow band or single-frequency mode. The potential for mode condensation depends on the relationship between frequency and stability for a given system. In general, the instability conditions for single-mode solutions of the GLE are well-known \cite{aranson_world_2002, garcia-morales_complex_2012}. The calculation has also been repeated in the context of an FEL \cite{ng_ginzburg-landau_1998}, and since the form of the chorus GLE is the same as the FEL GLE, the results can be easily translated. As with FELs, there are two types of instabilities. The Benjamin-Feir instability occurs when all single modes are unstable \cite{benjamin1967disintegration}. On the other hand, the Eckhaus instability \cite{eckhaus_studies_1965} occurs when a mode is surrounded by a band of stable modes but is unstable outside of this band. Both instability conditions are given by inequalities. 

In this work, we apply insights from the study of single-mode operation in FELs to the case of magnetospheric chorus. First, we introduce the GLE and its single-mode solutions. We then outline the process for obtaining the inequalities that express the linear stability conditions for these solutions. Next, we introduce a perturbation approximation for the linear growth rate of the whistler, which is one of the key physical parameters which determines the stability behavior. From this approximation, we show that the model predicts no Benjamin-Feir instability. Again employing the perturbation approximation for the linear growth rate, we obtain a simple expression for the Eckhaus instability condition, which allows one to compute the width of the stable frequency band. Importantly, we find that both instability conditions are approximately independent of the specific parameters of the system insofar as the basic assumptions of the FEL model are met, including that the resonance frequency must be near the maximum linear growth rate frequency. Last, we perform numerical simulations to support these analyses, and demonstrate the occurrence of mode condensation . 

In short, this study provides a new perspective on the development of narrow band chorus from a noisy spectrum through the process of mode condensation. This can be viewed as the tendency for modes outside the Eckhaus stability band to transition to modes within the stability band. In addition, this work shows that, under suitable assumptions, the stability conditions and bandwidth estimates can be reduced to a simple form. Ultimately, these results provide another interesting example of the insights which can be gained into the problem of magnetospheric chorus by drawing upon the FEL model and existing FEL literature. 

\section{Overview of the FEL Model}

The foundational equations of the FEL model (equations (7) - (9) in \citeA{soto-chavez_chorus_2012}) describe the interaction of a monochromatic whistler wave with $N$ resonant electrons near the geomagnetic equator. The whistler is assumed to have the form $\bm B_w = B_w(t) (\cos \varphi, \sin \varphi,0)$, where $B_w(t) = B_w$ is the amplitude,  $\varphi = \omega t - k z+\phi(t)$ is the overall phase, the relationship $\omega(k)$ is given by the cold whistler dispersion relation, the $z$ direction is positive along the background magnetic field, and $\phi(t)=\phi$ is a time-dependent phase shift. Using Maxwell's equations under a slowly varying amplitude and phase approximation, $B_w$ and $\phi$ are combined into a single equation. The electrons are coupled to the whistler by a current in Ampere's law. Importantly, in this work we focus on dynamics at the geomagnetic equator, where background field inhomogeneities are ignored, and utilize the background field $\bm B_0 = B_0 \hat z$. Under these magnetospheric conditions, the electron velocity perpendicular to the background field can assumed to be constant, hence $v_\perp = v_{\perp 0}$ \cite{Matsumoto_Cluster_1981,vomvoridis1979test}. In addition, the electrons are assumed to initially be monoenergetic, so their initial velocity parallel to the background field is simply $v_{z0}$. 

Under the above assumptions, one obtains a set of 2$N$+2 equations of the same form as the FEL equations. It was shown by \cite{soto-chavez_chorus_2012} that this multiplicity of equations can be approximated in the linear regime by a set of just three first-order linear differential equations in terms of collective variables \cite{bonifacio_collective_1986}. More recently, \citeA{bonham_whistler_2025} derived the corresponding nonlinear collective variable equations for the system (their equations (15) - (17)), allowing for the analytical investigation of nonlinear effects such as amplitude saturation and post-saturation amplitude modulations. As established in the context of FELs by \citeA{ng_ginzburg-landau_1998}, the single-mode FEL model can be extended to accommodate the behavior of a wave packet with multiple frequencies and an amplitude that varies in both time and space. This basic procedure for the extension involves using Fourier analysis to convert frequencies to time derivatives, and Taylor expanding the linear growth rate about a reference frequency $\omega_s$ near the principal mode. The resulting equation for the chorus wave packet is the cubic complex Ginzburg-Landau equation, one of the most celebrated nonlinear equations in physics \cite{aranson_world_2002}. 

\section{The Ginzburg-Landau Equation}

\subsection{Ginzburg-Landau Equation in Standard Form}

To begin, we briefly review the GLE for the evolution of chorus due to resonant interactions with radiation belt electrons \cite{bonham_whistler_2025},
\begin{equation} \label{Eqn: GLE}
    \frac{\partial \Phi}{\partial \zeta} = \Phi + (1+ic_1)\frac{\partial^2 \Phi}{\partial \tau^2} -(1+ic_2)|\Phi|^2\Phi,
\end{equation}
where $c_1 = -\alpha_r/\alpha_i$ and $c_2 = -\beta_r/\beta_i$. This form of the GLE, dependent on just two real constants, is the most simple and standard, making it well suited for analytical manipulation and comparison with existing literature \cite{aranson_world_2002}. The parameters $\alpha_r +i\alpha_i \equiv \alpha$ and $\beta_r + i \beta_i \equiv \beta$ are defined by, 
\begin{align}
    \alpha &\equiv \left. \frac{\partial^2 \lambda_0(\omega)}{\partial\omega^2} \right|_{\omega_s} \\
    \beta &\equiv \left. -2 u h_c \lambda_0 \left(  -\frac{1}{g\lambda_0} + \frac{ \delta}{g \lambda_0^2} + \frac{u h_c}{\lambda_0^4} + \frac{u h_c}{|\lambda_0|^4}\right) \right |_{\omega_s}. \label{Eqn: beta}
\end{align}
Physically, $\alpha$ is the dispersion of the linear growth rate curve of the whistler, $\lambda_0(\omega) =\lambda_0$. The linear growth rate is itself defined as the dominant root of the characteristic equation of the linear collective variable equations (equation (20) in \citeA{soto-chavez_chorus_2012}),
\begin{equation} \label{Eqn: Cubic Equation}
    \lambda_0^3+\delta\lambda_0^2+ug h_c = 0.
\end{equation}
The constants $\delta$, $u$, $g$, and $h_c$ are defined as, 
\begin{equation}
\begin{aligned}
    \delta &\equiv k (v_{z0}-v_r) \\
    u      &\equiv v_{\perp 0}/2 \\
    g      &\equiv \frac{s \Omega_{e0} \omega_{pr}^2 u}{k c^2} \\
    h_c    &\equiv -\frac{k}{\gamma_0}(1- \frac{\gamma_{\perp 0}^2\eta_{z0}^2}{\gamma_0^2 c^2} -\frac{\gamma_{\perp 0}^2\Omega_{e0}\eta_{z0}}{\gamma_0^2 c^2 k}),
\end{aligned}
\end{equation}
where $v_r \equiv (\omega-\Omega_{e0}/\gamma_0)/k$ is the gyroresonance velocity, $s\equiv \omega/(\Omega_{e0}-\omega)$ comes from the dispersion relation, $\Omega_{e0} \equiv eB_0/m_e$ is the electron cyclotron frequency due to the background geomagnetic field,  $\omega_{pr} \equiv (n_r e^2/\epsilon_0 m_e)^{1/2}$ is the resonant electron plasma frequency, $\gamma_0 \equiv (1 - \bm v_0^2/c^2)^{-1/2}$ is the resonant electron initial Lorentz factor, $\gamma_{\perp 0} \equiv (1 - v_{\perp 0}^2/c^2)^{-1/2}$ is the resonant electron perpendicular Lorentz factor, $\eta_{z 0} \equiv \gamma_0 v_{z0}$ is the resonant electron initial proper velocity parallel to $\bm B_0$. Lastly, $n_r$ is the number density of resonant electrons, $e$ is the elementary charge, $\epsilon_0$ is the permittivity of free space, $m_e$ is the mass of the electron, and $c$ is the speed of light in vacuum. Importantly, $\delta$, $g$, and $h_c$ all depend implicitly on $\omega$ through their dependence on $k$. 

In exchange for the simplicity of equation \eqref{Eqn: GLE}, one must apply two transformations to recover the physical field, $\Phi(\zeta,\tau) \rightarrow A(z,t) \rightarrow \bm{B}_w(z,t)$, first converting from the rescaled collective variable in scaled space and time coordinates, $\Phi(\zeta,\tau)$, to the standard collective variable in physical space and time coordinates, $A(z,t)$, and finally to the whistler magnetic field, $\bm B_w$. To convert from $\Phi(\zeta,\tau)$ to $A(z,t)$ we apply the transformations, 
\begin{align}
    A(z,t) &= \Phi_0 \Phi(\zeta,\tau)\exp i[K_0 \zeta + \Omega_0 \tau] \\
    \zeta &= z/z_0 \\
    \tau &= t/t_0+z/v_0t_0.
\end{align}
The scaling constants are given by, 
\begin{equation}
\begin{aligned}
    z_0 &= 1/(-\bar{\lambda}_i+\bar{\mu}_i^2/2\bar{\alpha}_i),& t_0^2 &= \bar{\alpha}_i z_0/2, \\
    v_0 &= 1/(-\bar{\mu}_r +\bar{\alpha}_r \bar{\mu}_i /\bar{\alpha}_i),& \Phi_0^2 &= 1/z_0\bar{\beta}_i, \\
    K_0 &= z_0(\bar{\lambda}_r - \bar{\alpha}_r\bar{\mu}_i^2/2\bar{\alpha}_i^2),& \Omega_0 &= \bar{\mu}_i t_0 / \bar{\alpha}_i,
\end{aligned}
\end{equation}
where $ \mu \equiv 1 + \left .\frac{\partial \lambda_0}{\partial\omega} \right|_{\omega_s}$, the subscripts denote real and imaginary parts, the overbars denote division by the group velocity, e.g. $ \bar \alpha_i \equiv \alpha_i /v_g$, and the group velocity can be calculated from the cold whistler dispersion relation. Finally, the whistler field may be recovered from $A$ via the relation, $A = (e/m_e)B_w e^{i(\phi-\delta t)}$. Or if we denote the complex phase of $A$ as $\phi_A$, then the whistler amplitude and phase is fully specified by,
\begin{equation}
\begin{aligned}
    B_w &= \frac{m_e}{e} |A| \\
    \phi &= \phi_A + \delta t.
\end{aligned}
\end{equation}
\subsection{Single-Mode Solutions}

It can be verified by direct substitution that the single-frequency mode, 
\begin{equation} \label{Eqn: SM Solution}
    \Phi = \sqrt{1-\omega_0^2} \exp i\left(k_0 \zeta-\omega_0 \tau + \phi_0\right)
\end{equation}
is a solution to equation \eqref{Eqn: GLE}, for any $\omega_0 \in (-1,1)$ where $k_0 = -c_2 - (c_1-c_2)\omega_0^2$. Note, the constant $\phi_0 \in \mathbb R$, which determines the phase at $\zeta = \tau = 0$ is hereafter assumed zero for simplicity. To extract the whistler frequency corresponding to this solution, we transform from $\Phi(\zeta,\tau)$ to $A(z,t)$, then employ the operator identity $i \partial/\partial t = \Delta\omega \equiv (\omega -\omega_s)$ which originates from the Fourier analysis which led to the GLE. Applying this operator to $A$ yields, 
\begin{equation} \label{Eqn: SM Frequency Long}
    \omega = \omega_s - \frac{\mu_i}{\alpha_i} + \frac{\omega_0}{t_0}.
\end{equation}
By Taylor expansion of $\lambda_i(\omega)$ it can be shown that the frequency with the maximum linear growth rate, $\omega_m$, may be written as $\omega_m = \omega_s - \mu_i/\alpha_i + O\left(|\omega_m - \omega_s|^2\right)$ \cite{ng_ginzburg-landau_1998}. Since the GLE derivation requires that the reference frequency, $\omega_s$, is near the maximum linear growth rate frequency, $\omega_m$, the error is small by construction. For example, for typical magnetospheric parameters \cite{Omura2008Theory,lampe2010nonlinear,santolik_spatio-temporal_2003} and $\omega_s=\omega_\delta$ (where $\omega_\delta$ is defined in section \ref{Sec: Stability Bandwidth}) the relative error is roughly one part in a thousand. Therefore, the frequency of the single mode solution is approximately a shift of $\omega_0/t_0$ away from the maximum growth mode and equation \eqref{Eqn: SM Frequency Long} may be written,  
\begin{equation}
    \omega \simeq \omega_m + \frac{\omega_0}{t_0}.
\end{equation}
\subsection{Instability Criteria}

The linear stability of equation \eqref{Eqn: SM Solution} can be analyzed by adding a small perturbation to $\Phi$, then substituting the perturbed expression into the GLE, then re-solving the GLE to linear order. If the perturbed solution grows exponentially, then the original solution is linearly unstable. This procedure, sketched here, is well known in general, and has also been conducted in particular in the FEL context. When written in standard form, the FEL GLE is identical to the chorus GLE, apart from the numerical values of $c_1$ and $c_2$. Thus, the stability results can be carried over straightforwardly. Here we sketch the analysis, presented for FELs by \citeA{ng_ginzburg-landau_1998}, and apply it to the problem of magnetospheric chorus. 

It will be shown from the instability conditions (equations \eqref{Eqn: BF Criterion} and \eqref{Eqn: Eckhaus Criterion}) that if the mode with the maximum linear growth rate (the $\omega_0 = 0$ mode) is unstable then all single modes are unstable, corresponding to the Benjamin-Feir instability. We first sketch this case. To analyze the linear stability of the mode, $\Phi = \exp (-ic_2\zeta)$, we add small perturbations $a(\zeta,\tau) = \tilde{a}(\zeta) \cos(\omega_d \tau)$ and $\theta(\zeta,\tau) = \tilde{\theta}(\zeta) \cos(\omega_d \tau)$ to the amplitude and phase respectively to obtain the perturbed mode $\tilde \Phi$, 
\begin{equation}
    \Phi \rightarrow \tilde{\Phi} = \bigl(1+\tilde{a}(\zeta) \cos(\omega_d \tau) \bigr )\exp{i\left[-c_2 \zeta +\tilde{\theta}(\zeta) \cos(\omega_d \tau)\right]}.
\end{equation}
Using this ansatz in the GLE and linearizing with respect to  $\tilde a(\zeta)$ and $\tilde \theta(\zeta)$ yields the exponential solutions $\tilde a(\zeta) = \tilde a_0\exp(\Lambda \zeta)$ and $\tilde \theta(\zeta) = \tilde \theta_0 \exp (\Lambda \zeta)$, where $\tilde a_0 \equiv \tilde a(0)$,  $\tilde \theta_0 \equiv \tilde \theta(0) $, and the growth rate of the perturbations is, 
\begin{equation}
    \Lambda = -(1+\omega_d^2) \pm \sqrt{1-2 c_1 c_2 \omega_d- c_1^2 \omega_d^4}.
\end{equation}
If $\Lambda$ has a positive real part then the perturbations will grow exponentially and $\Phi$ will be unstable. This yields the Benjamin-Feir-Newell \textit{instability} criterion, 
\begin{equation} \label{Eqn: BF Criterion}
    1 + c_1 c_2 < 0. 
\end{equation}
If the $\omega_0=0$ mode is stable, then there is a band of stable values of $\omega_0$ outside of which all single modes are unstable, known as the Eckhaus instability. To show this, we introduce the aforementioned perturbations to the $\omega_0 \not = 0$ solution, 
\begin{equation} \label{Eqn: Perturbed SM}
    \Phi \rightarrow \tilde{\Phi} = 
    \sqrt{1-\omega_0^2}\Big(1+\tilde{a}(\zeta) \cos(\omega_d \tau) \Big )
    \exp{i\left[k_0 \zeta - \omega_0 \tau +\tilde{\theta}(\zeta) \cos(\omega_d \tau)\right]}.
\end{equation}
As before, linearizing the GLE using this ansatz yields exponential solutions for $\tilde a(\zeta)$ and $\tilde \theta(\zeta)$. The growth rate of low frequency perturbations ($\omega_d \ll 1)$ is given by, 
\begin{equation}
    \Lambda \simeq \left\{
        \begin{gathered}
        -2(1-\omega_{0}^{2}), \\
        -2\left[1 + c_1 c_2 + \frac{2 \omega_0^2}{\omega_0^2 - 1}(1+c_2^2)\right] \frac{\omega_d^2}{2}
        +2 i \omega_0 \omega_d (c_2 - c_1).
        \end{gathered}
    \right.
\end{equation}
If $\Lambda$ has a positive real part then $\Phi$ will be linearly unstable. This yields the Eckhaus \textit{instability} criterion,
\begin{equation} \label{Eqn: Eckhaus Criterion}
    1 + c_1 c_2 + \frac{2 \omega_0^2}{\omega_0^2 - 1}(1+c_2^2) < 0.
\end{equation}
Notice that for $\omega_0 = 0$ this reduces to Eqn. \eqref{Eqn: BF Criterion}. Also notice that since in general $0 < \omega_0^2 < 1$, the third term on the left-hand side is always negative. Hence, if the $\omega_0 = 0$ mode is unstable, then all modes are unstable. Conversely, if the $\omega_0 = 0$ mode is stable, then there will always be some critical value $|\omega_0| > \omega_c$ above which the Eckhaus instability will always occur, since $\omega_0^2 -1$ diverges at $\omega_0 =\pm 1$. We calculate this critical frequency in the next section. 

\section{Stability Bandwidth and Reduced Instability Criteria} \label{Sec: Stability Bandwidth}

In this section, we simplify the instability criteria then use the result to show that there is no Benjamin-Feir instability and calculate the range of frequencies included in the Eckhaus stability band. Notice both instability criteria (equations \eqref{Eqn: BF Criterion} and \eqref{Eqn: Eckhaus Criterion}) depend solely on the constants $c_1$ and $c_2$, 
\begin{align}
    c_1 &\equiv - \left. \frac{\alpha_r}{\alpha_i} \right|_{\omega_s} \equiv \frac{\lambda_r''(\omega_s)}{\lambda_i''(\omega_s)} \\
    c_2 & \equiv - \left.\frac{\beta_r}{\beta_i} \right|_{\omega_s},
\end{align}
which must be evaluated at some reference frequency $\omega_s$ near the maximum of the linear growth rate $\omega_m$. Ideally, one would express the instability criteria in terms of the input parameters of the system ($v_{\perp 0}$, $\gamma_0$, etc.), rather than the more compact but less physical coefficients $c_1$ and $c_2$. However, $c_1$ and $c_2$ depend on $\lambda_0$, which is the solution to a cubic equation, so that the exact expressions for $\lambda_0$ and $\lambda_0''$ are unwieldy. 

One can simplify the analysis by taking the reference frequency $\omega_s$ to be the gyroresonance frequency, i.e. the frequency at which $\delta = 0$. Since this resonance occurs near a local maxima of $\lambda_0$, this satisfies the assumption that $\omega_s$ is near $\omega_m$. We denote the $\delta=0$ frequency by $\omega_\delta$. Now the expression for $\lambda_0(\omega_\delta)$ can be readily obtained by setting $\delta=0$ in equation \eqref{Eqn: Cubic Equation}, which yields, 
\begin{equation} \label{Eqn: lambda appx short}
    \lambda_0(\omega_\delta) = (-u g h_c)^{1/3} .
\end{equation}
Equation \eqref{Eqn: lambda appx short} is exact at $\delta=0$, but the behavior at nearby points is not accurate enough to properly determine its second derivative, $\alpha$. However, by a perturbation expansion of the cubic equation, one can obtain an order-by-order approximation for $\lambda_0(\omega)$ in powers of $\delta$ which is valid near $\omega_\delta$. By assumption, $\delta$ is small for resonant interactions, therefore the expansion can be used to obtain an excellent approximation for $\alpha$. The perturbation assumption $\lambda_0 = f_0+f_1\delta+f_2\delta+\cdots \ $ in equation \eqref{Eqn: Cubic Equation} yields, 
\begin{equation}
    (f_0^3+ u g h_c)+(f_0^2 + 3 f_0^2f_1)\delta + (2 f_0 f_1 + 3 f_0 f_1^2 + 3 f_0^2 f_2) \delta^2 + \dots =0.
\end{equation}
Solving order by order in $\delta$, one obtains the first three terms, 
\begin{equation} \label{Eqn: lambda appx}
    \lambda_0 \simeq (-u g h_c)^{1/3} - \frac{1}{3} \delta + \frac{1}{9}(- u g h_c)^{-1/3} \delta^2.
\end{equation}
Therefore, 
\begin{equation}
\begin{aligned}
    \frac{\partial^2 \lambda_0}{\partial \omega^2}
    &\simeq
    \frac{\partial^2 (-u g h_c)^{1/3}}{\partial \omega^2}
    -\frac{1}{3}\frac{\partial^2 \delta}{\partial \omega^2} \\
    & \quad + \frac{1}{9}\Bigg[
    \frac{\partial^2 (-u g h_c)^{-1/3}}{\partial\omega} \delta^2 +
    4 \frac{\partial (-u g h_c)^{-1/3}}{\partial \omega} \frac{\partial \delta}{\partial \omega} \delta
    \\
    & \quad + 2  (-u g h_c)^{-1/3} \left( \frac{\partial \delta}{\partial \omega} \right)^2
    + 2 (- u g h_c)^{-1/3} \delta \frac{\partial^2 \delta}{\partial \omega ^2}
    \Bigg].
\end{aligned}
\end{equation}
By evaluating at $\omega_\delta$, one has,  
\begin{equation}
    \alpha \simeq 
    \frac{\partial^2 (-u g h_c)^{1/3}}{\partial \omega^2}
    -\frac{1}{3}\frac{\partial^2 \delta}{\partial \omega^2} 
    + \frac{2}{9}(-u g h_c)^{-1/3} \left( \frac{\partial \delta}{\partial \omega} \right)^2.
\end{equation}
Finally, since $\delta$ may be written as, $\delta = \Omega_{e0}/\gamma_0 - \omega +k v_{z0}$, one obtains, 
\begin{equation} \label{Eqn: alpha appx}
    \alpha \simeq \frac{2}{9} (-u g h_c)^{-1/3} \left(\frac{v_{z0}}{v_g}-1 \right)^2,
\end{equation}
where $v_g >0$, and the subdominant contributions originating from the first two terms in \eqref{Eqn: lambda appx} have been dropped (these terms are insufficient to produce parabolic behavior in $\lambda_0$ near $\omega_\delta$, thus do not contribute meaningfully to the second derivative). Importantly, one must always take the root of $\lambda_0$ with the highest linear growth rate, i.e. the largest negative imaginary part, which corresponds to the branches $(-u g h_c)^{1/3} = |u g h_c|^{1/3}e^{-2 \pi i /3}$ and $(-u g h_c)^{-1/3} = |u g h_c|^{-1/3}e^{2 \pi i /3}$ in the above expressions. 

From equation \eqref{Eqn: alpha appx} one can readily show that, independent of the initial conditions, 
\begin{equation} \label{Eqn: c1 appx}
    c_1 \simeq \frac{1}{\sqrt{3}}.
\end{equation}
Using equation \eqref{Eqn: lambda appx} in equation \eqref{Eqn: beta} one has,
\begin{equation}
    \beta = 2 \sqrt{3}\frac{ u h_c}{g} e^{-i\pi/6}. 
\end{equation}
Thus $c_2$ may be written as, 
\begin{equation} \label{Eqn: c2 appx}
    c_2 = \sqrt{3}.
\end{equation}
From these expressions for $c_1$ and $c_2$, it is evident that the Benjamin-Feir instability criterion (eqn. \eqref{Eqn: BF Criterion}) cannot be met, hence the Benjamin-Feir instability cannot occur, irrespective of the system parameters. This means the mode with the maximum linear growth rate is always stable and surrounded by a band of stability. 

Furthermore, the Eckhaus instability criterion (eqn. \eqref{Eqn: Eckhaus Criterion}) becomes $\omega_0^2>1/5$, so that the critical onset frequency is $\omega_c = 1/\sqrt5$. Thus, the region of stability is $-\omega_c<\omega_0<\omega_c$, or approximately $-.45<\omega_0<.45$. This corresponds to a stability bandwidth for the whistler of $\Delta\omega \simeq 2\omega_c/t_0$. For typical magnetospheric parameters \cite{bonham_whistler_2025,soto-chavez_chorus_2012}, this is approximately $\Delta \omega \simeq 0.02 \, \Omega_{e0}$. This quantity refers to the bandwidth of potentially stable single modes, rather than the bandwidth of a wavepacket. Although the physical interpretation differs, the value is the of the same order as CLUSTER spacecraft observations of chorus bandwidths, which are typically $0.03 \,\Omega_{e0} $ to $ 0.2 \, \Omega_{e0}$ \cite{santolik_frequencies_2008}. 

\section{Numerical Simulations}

We now demonstrate the behavior of single-mode solutions within and outside of the Eckhaus stability band derived above. We solve the GLE numerically using Mathematica's standard numerical differential equation tools, employing the method of lines option to convert the PDE to an initial value problem in the continuous variable $\zeta$ by discretizing the variable $\tau$ on a grid. This results in an ODE in $\zeta$ at each grid point of $\tau$, which we integrate using the explicit modified midpoint method option. Commensurate with our periodic boundary conditions in time, the $\tau$ derivatives are computed using the pseudospectral option. To incorporate noise while preserving the periodic boundary conditions, we add 256 harmonics of a reference frequency, $\omega_0$, with random small amplitudes drawn from a uniform distribution. In all cases, we take $c_1=1/\sqrt3$ and $c_2=\sqrt3$, focusing on the magnetospheric conditions associated with the FEL model. 

We present three characteristic behaviors. First, single modes within the Eckhaus stability band persist under large amplitude perturbations. Second, single modes outside the Eckhaus stability band may propagate temporarily, but eventually evolve towards a stable single mode. This transition process can occur within a few wavelengths, or can occur in a more distributed manner, presenting here as sweeping crescents in the spacetime diagrams. Last, an initially noisy spectrum of modes can condense to a single-mode state. 

\begin{figure}[!ht] 
    \centering
    \includegraphics[width=.9\textwidth]{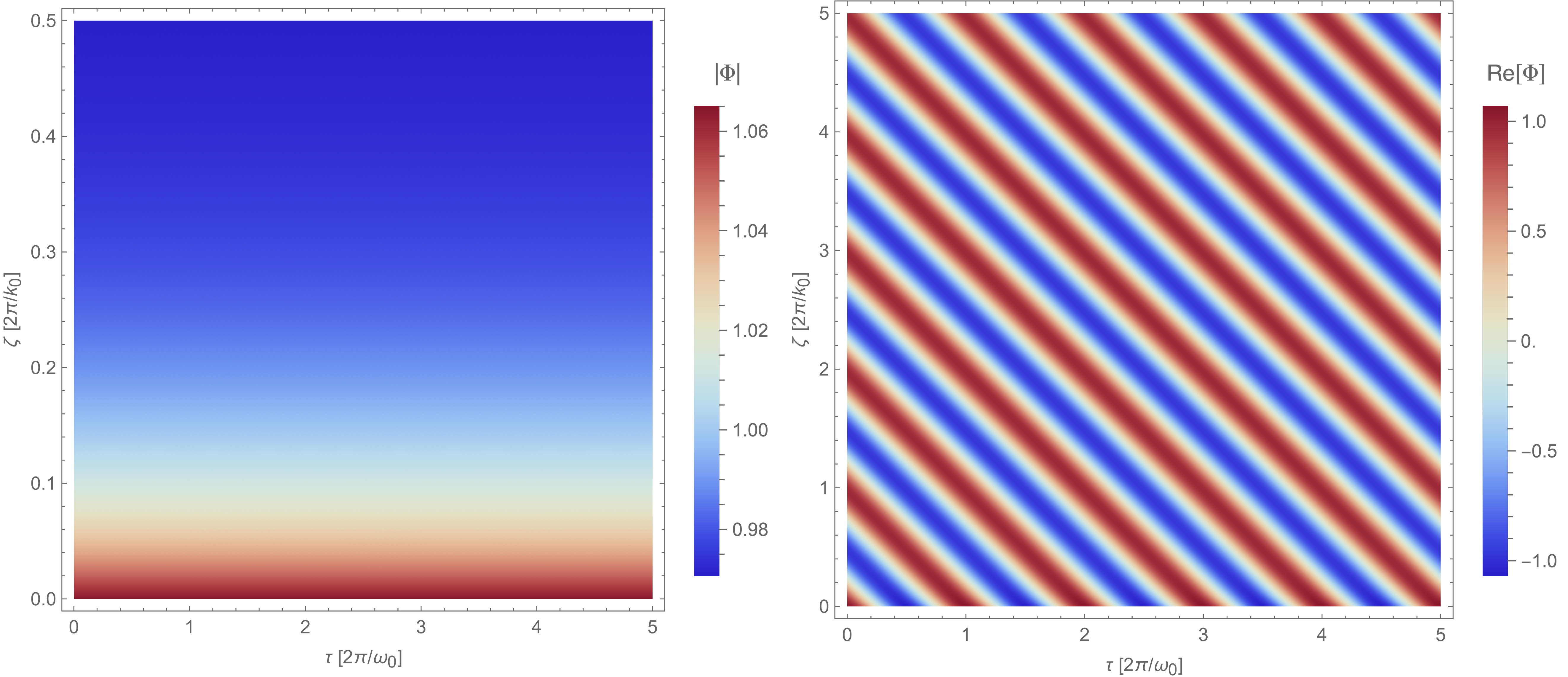}
    \caption{Spacetime plots of the amplitude (left) and real part (right) of $\Phi$, demonstrating that a stable mode ($\omega_0=1/4$) dissipates a $10\%$ amplitude perturbation within one wavelength.} 
    \label{Fig 1: Stable w_0}
\end{figure}

We begin by demonstrating the behavior of single-mode solutions within the Eckhaus stability band, choosing $\omega_0=1/4$ as a representative mode and including a moderate amplitude perturbation. Hence, the initial condition $\Phi(\zeta=0) = (1+\tfrac{1}{10})(1-\omega_0^2)^{1/2}\exp{(-i\omega_0\tau)}$. In the absence of the amplitude perturbation, this mode is an exact solution of the GLE, and will persist unchanged. Figure \ref{Fig 1: Stable w_0} shows that such steady state propagation occurs even when the amplitude is initially perturbed. As can be seen in the left panel, the amplitude perturbation decays within a fraction of a wavelength to $|\Phi|=0.97$, or the amplitude which is self-consistent with the single mode frequency-amplitude relation, $|\Phi|=(1-\omega_0^2)^{1/2}$. Thereafter, although this it is not shown in the figure, the mode persists indefinitely. 

\begin{figure}[!ht] 
    \centering
    \includegraphics[width=.9\textwidth]{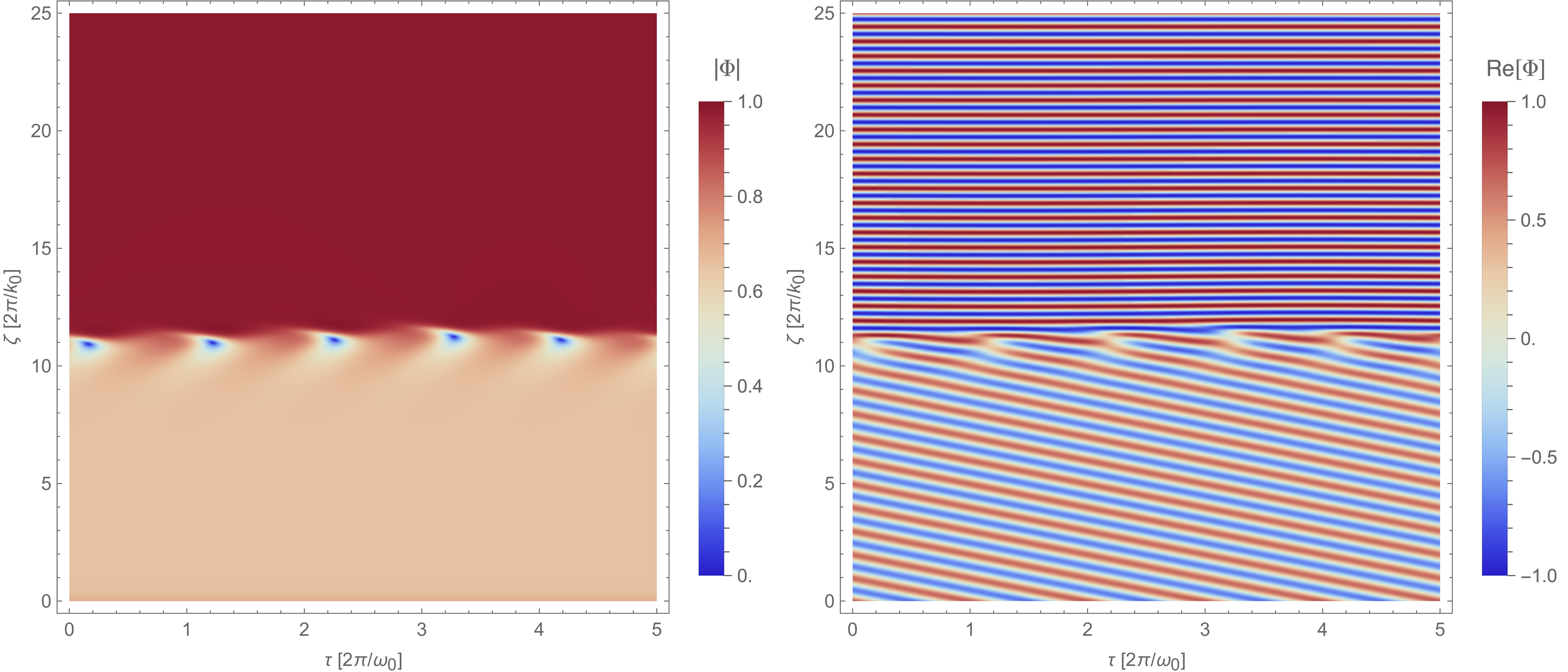}
    \caption{An unstable mode ($\omega_0=3/4$) dissipates an initial $10\%$ amplitude perturbation within one wavelength. The resulting mode, still unstable, persists for several wavelengths, then transitions to a stable mode.} 
    \label{Fig 2: Unstable w_0}
\end{figure}

Next, we consider the behavior of modes outside the stability band. Figure \ref{Fig 2: Unstable w_0} presents spacetime plots for the same initial conditions as in figure \ref{Fig 1: Stable w_0}, but with the unstable frequency $\omega_0=3/4$. So that the eventual transition to stability is not driven merely by numerical errors, in addition to the initial amplitude perturbation we include small overall constant. Hence the initial condition is, $\Phi(\zeta=0) = (1+\tfrac{1}{10})(1-\omega_0^2)^{1/2}\exp{(-i\omega_0\tau)} + 10^{-9}$. As with the previous case, the amplitude perturbation decays quickly, in a fraction of a wavelength, such that the amplitude becomes self-consistent with the frequency, $|\Phi| = (1-\omega_0^2)^{1/2} = 0.66$. The resultant mode, although unstable, is still a solution to the GLE, consistent with the fact that the mode persists for several wavelengths before defects and phase ripples appear which quickly force the system towards a stable mode. 

\begin{figure}[!ht]
    \centering
    \includegraphics[width=.9\textwidth]{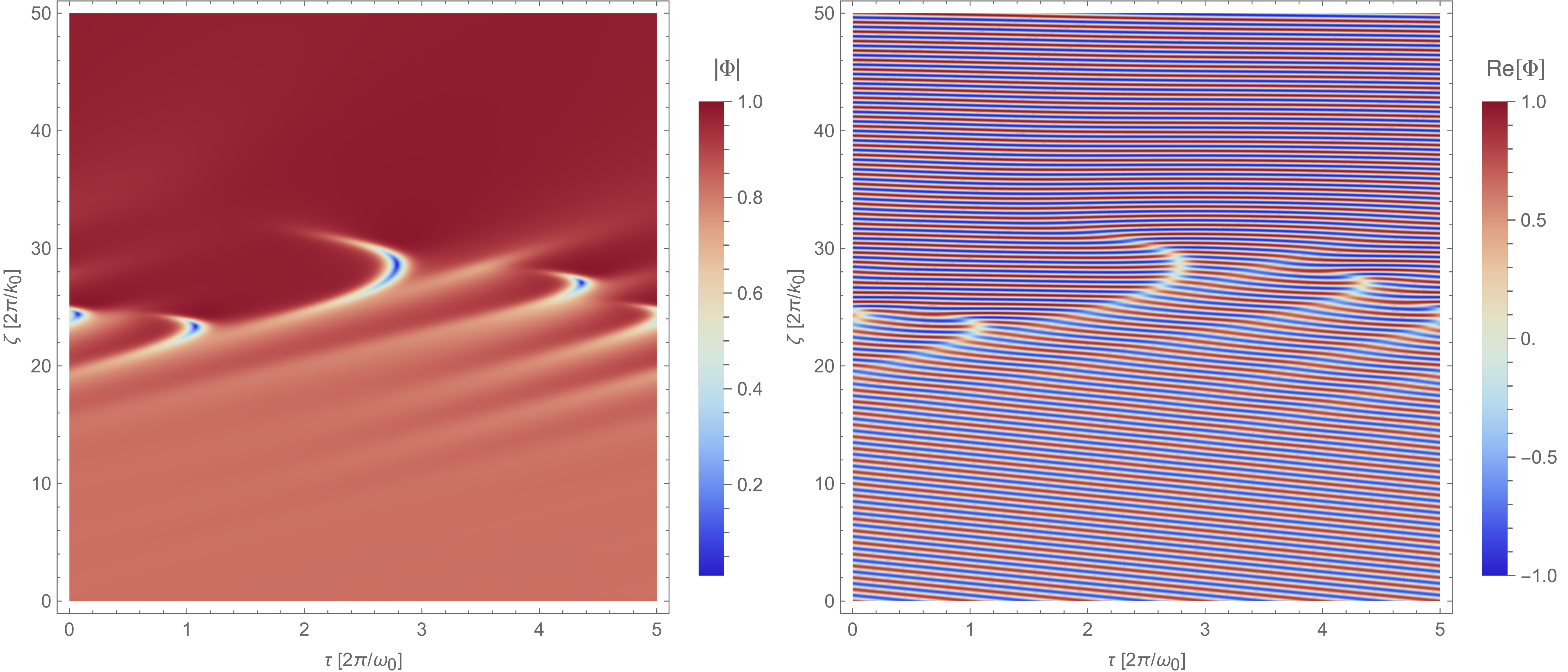}
    \caption{An unstable mode ($\omega_0=0.55$) nearer to the stability threshold gradually transitions to a stable mode due to the presence of noise.} 
    \label{Fig 3: Unstable w_0 + Noise}
\end{figure}

Having demonstrated the rapid dissipation of amplitude perturbations and the tendency of unstable modes to evolve towards stable modes, we now consider the effects of noise on a mode closer to the stability threshold. In figure \ref{Fig 3: Unstable w_0 + Noise} we consider the initial condition $\Phi(\zeta=0) = (1-\omega_0^2)^{1/2}\exp{(-i\omega_0\tau)} + \delta\Phi$, where $\omega_0=0.55$, and $\delta\Phi$ includes harmonics of $\omega_0$ with random amplitudes drawn uniformly from $|\delta\Phi| \in \left[0,(1-\omega_0)^2/10^3\right]$. In the absence of the noise, since the mode is an exact solution of the GLE, one would need to rely on numerical errors to stimulate the onset of the stability transition. Here this onset is due to the deliberate noise, so the irregularities appear within a few wavelengths of the origin. However, since this mode is closer to the stability threshold than that of figure \ref{Fig 2: Unstable w_0}, as dictated by equation \eqref{Eqn: lambda appx} the instabilities which drive the system toward a stable state have a slower growth rate. Rather than occurring over a single wavelength, the stability transition is distributed across tens of wavelengths, presenting as sweeping crescents in the spacetime diagrams. 

\begin{figure}[!ht]
    \centering
    \includegraphics[width=.9\textwidth]{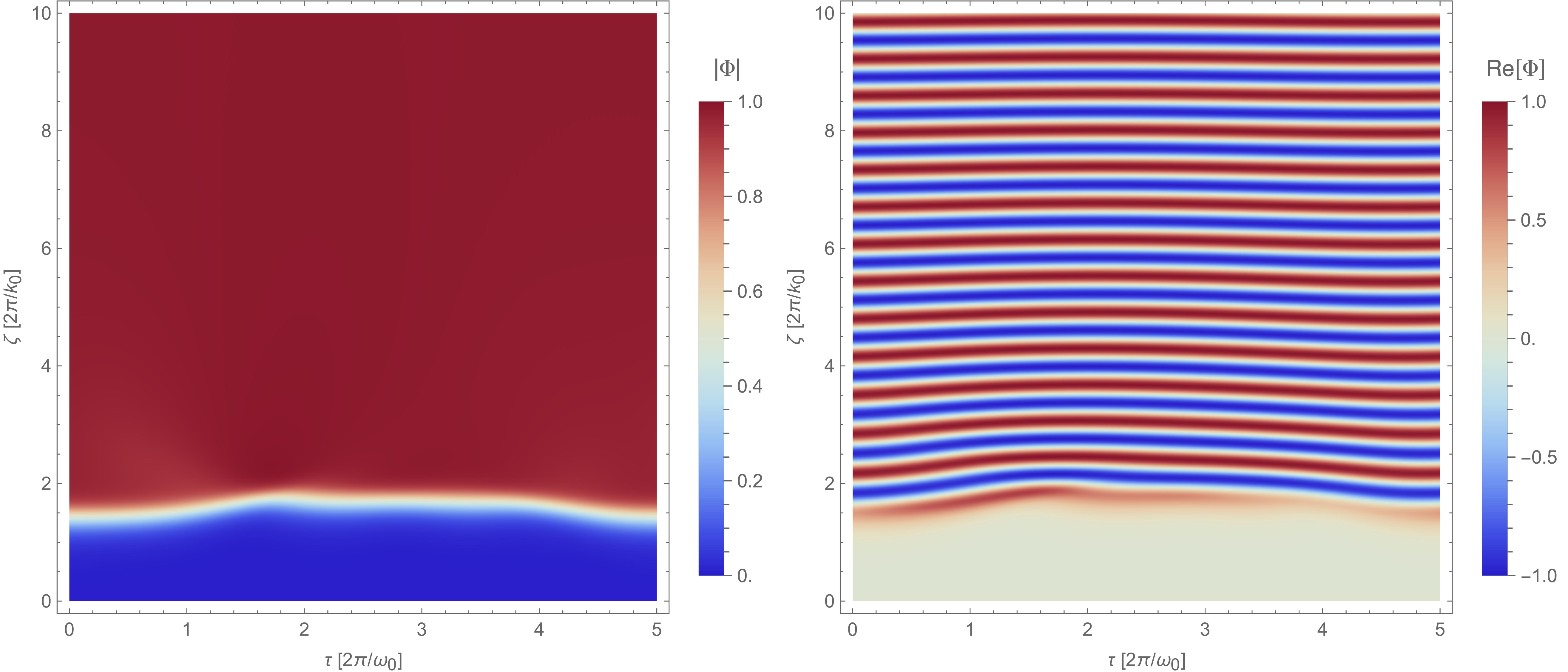}
    \caption{Condensation of 256 modes with small random amplitudes down to a single stable mode. }
    \label{Fig 4: Mode Condensation}
\end{figure}

Finally, in figure \ref{Fig 4: Mode Condensation} we consider an initial condition consisting solely of a noisy spectrum of modes with small random amplitudes. The noise consists of harmonics of the reference frequency $\omega_0=3/4$ with random amplitudes drawn uniformly from the range $\delta\Phi_m\in\left[0,(1-\omega_0^2)^{1/2}/ 10^{4}\right]$, so that the initial condition is, $\Phi = \sum_{m=-128}^{128}|\delta\Phi_m| e^{-im\omega_0\tau/N_\tau}$, where $N_\tau=5$ is the number of periods included in the range. The broad spectrum of modes present in the initial state quickly transfer their energy to the mode with the highest linear growth rate. This behavior, known as mode condensation, has been investigated in the context of FELs as a potential mechanism for obtaining beam purity. However, it is not the only potential behavior. Both Benjamin-Feir and Eckhaus unstable systems have been shown numerically to exhibit phase turbulence \cite{garcia-morales_complex_2012, ng_ginzburg-landau_1998}. However, since these behaviors depend strongly on $c_1$ and $c_2$, which we have shown to be near particular values for the resonant interactions described by the FEL model, here we do not expect phase turbulence to occur. 

\section{Discussion \& Conclusion}

In this study, we investigated the stability of single-mode chorus propagation in the magnetosphere using the Ginzburg-Landau equation. This equation, recently proposed as a nonlinear description for whistler-mode chorus, originates from the FEL model \cite{bonham_whistler_2025, soto-chavez_chorus_2012}. By drawing upon insights from the FEL literature, we found that the single mode with the highest linear growth is stable and surrounded by a narrow band of stability, outside of which modes are vulnerable to the Eckhaus instability. After briefly reviewing the linear stability analysis, by a perturbation expansion of the linear growth rate, we found that the width of the Eckhaus stability band is approximately two percent of the background electron cyclotron frequency. We then conducted a brief numerical study to illustrate three key behaviors - the persistence of stable frequencies under amplitude perturbations, the evolution of unstable frequencies towards stable ones, and the condensation of a spectrum of random modes down to a single mode. 

Finally, we note an additional implication of this work of interest to the theory of chorus. From equation \eqref{Eqn: SM Solution} it can be readily seen that the single-mode solutions to the GLE contain an explicit self-consistent coupling between the amplitude and frequency of the mode, namely $|\Phi| =(1-\omega_0^2)^{1/2}$. Thus the amplitude evolution that occurs for Eckhaus unstable modes, as seen in figures \ref{Fig 2: Unstable w_0} and \ref{Fig 3: Unstable w_0 + Noise}, implies frequency evolution. Furthermore, distinct frequencies can be seen in the right panels of those figures, where the contours of the real part of the mode undergo a change in angle before and after the stability transition. Because the initial value of $\omega_0$ may be above or below the stability band, the frequency evolution may be downwards (if $\omega_0>\omega_c$) or upwards (if $\omega_0<-\omega_c$) as the mode transitions from an unstable mode into the stable band. The implications for magnetospheric chorus, including the frequency chirp rate and duration, will be the subject of a forthcoming publication. 


\section*{Conflict of Interest}

The authors declare there are no conflicts of interest for this manuscript.

\section*{Data Availability Statement}

Data were not used, nor created for this research. All figures were produced with Wolfram Mathematica~14.2 using the built-in numerical differential equation solver. 


\acknowledgments
The authors acknowledge support from the NSF Award No. 2209471.  

\bibliography{references}

\end{document}